\documentclass[conference]{IEEEtran}
\IEEEoverridecommandlockouts
\usepackage{algorithmic}
\usepackage{amsmath,amssymb,amsfonts}
\usepackage{cite}
\usepackage[acronym]{glossaries}
\usepackage{graphicx}
\usepackage{mathrsfs}
\usepackage{subcaption}
\usepackage{textcomp}
\usepackage{xcolor}
\def\BibTeX{{\rm B\kern-.05em{\sc i\kern-.025em b}\kern-.08em
    T\kern-.1667em\lower.7ex\hbox{E}\kern-.125emX}}
\usepackage{hyperref}
\usepackage{algorithm2e}

\newacronym{ac}{AC}{alternating current}%
\newacronym{bnb}{B\&B}{branch-and-bound}%
\newacronym{mv}{MV}{medium voltage}%
\newacronym{hv}{HV}{high voltage}%
\newacronym[\glsshortpluralkey={DAPs}, \glslongpluralkey={day-ahead prices}]{dap}{DAP}{day-ahead prices}%
\newacronym[\glsshortpluralkey={DoFs}, \glslongpluralkey={degrees of freedom}]{dof}{DoF}{degree of freedom}%
\newacronym[\glsshortpluralkey={BZNs}, \glslongpluralkey={bidding zones}]{bzn}{BZN}{bidding zone}%
\newacronym[\glsshortpluralkey={MV-Ss}, \glslongpluralkey={MV-substations}]{mvs}{MV-S}{MV-substation}%
\newacronym[\glsshortpluralkey={HV-Ss}, \glslongpluralkey={HV-substations}]{hvs}{HV-S}{HV-substation}%
\newacronym[\glsshortpluralkey={CBs}, \glslongpluralkey={cycle bases}]{cb}{CB}{cycle basis}%
\newacronym{fsc}{FSC}{fundamental system of cycles}%
\newacronym[\glsshortpluralkey={CEs}, \glslongpluralkey={cycle-edges}]{ce}{CE}{cycle-edge}%
\newacronym[\glsshortpluralkey={UFCEs}, \glslongpluralkey={unique fundamental cycle edges}]{ufce}{UFCE}{unique fundamental cycle edge}%
\newacronym[\glsshortpluralkey={UESs}, \glslongpluralkey={unique edge sets}]{ues}{UES}{unique edge set}%
\newacronym[\glsshortpluralkey={rUESs}, \glslongpluralkey={reduced unique edge sets}]{rues}{rUES}{reduced unique edge set}%
\newacronym{dsr}{DSR}{distribution system reconfiguration}%
\newacronym{rdsr}{R-DSR}{restrictable distribution system reconfiguration}%
\newacronym{cdsr}{C-DSR}{complete distribution system reconfiguration}%
\newacronym{rrdsr}{RR-DSR}{relaxed radiality distribution system reconfiguration}%
\newacronym{hdsr}{H-DSR}{heuristic distribution system reconfiguration}%
\newacronym[\glsshortpluralkey={FCs}, \glslongpluralkey={fundamental cycles}]{fc}{FC}{fundamental cycle}%
\newacronym{mip}{MIP}{mixed-integer program(-ming)}%
\newacronym[\glsshortpluralkey={MILPs}, \glslongpluralkey={mixed-integer linear programs}]{milp}{MILP}{mixed-integer linear program(-ming)}%
\newacronym[\glsshortpluralkey={MIQPs}, \glslongpluralkey={mixed-integer quadratic programs}]{miqp}{MIQP}{mixed-integer quadratic program}%
\newacronym[\glsshortpluralkey={MISOCs}, \glslongpluralkey={mixed-integer second-order cone programs}]{misoc}{MISOC}{mixed-integer second-order cone program}%
\newacronym[\glsshortpluralkey={MINLPs}, \glslongpluralkey={mixed-integer non-linear programs}]{minlp}{MINLP}{mixed-integer non-linear program}%
\newacronym{nlp}{NLP}{non-linear program}%
\newacronym{nlpr}{NLP-R}{NLP relaxation}%
\newacronym[\glsshortpluralkey={N-FPs}, \glslongpluralkey={nodal flexibility potentials}]{nfp}{N-FP}{nodal flexibility potential}%
\newacronym[\glsshortpluralkey={LI-FPs}, \glslongpluralkey={location-invariant flexibility potentials}]{lifp}{LI-FP}{location-invariant flexibility potential}%
\newacronym{lns}{LNS}{large neighborhood search}%
\newacronym{mcb}{MCB}{minimum cycle basis}%
\newacronym{pf}{PF}{power flow}%
\newacronym{acpf}{ACPF}{alternating current power flow}%
\newacronym{saidi}{SAIDI}{system average interruption duration index}
\newacronym{soc}{SOC}{second-order cone}%
\newacronym{dso}{DSO}{distribution system operator}%
\newacronym{dsoe}{DSO-E}{DSO Entity}%
\newacronym{entsoe}{ENTSO-E}{European Network of Transmission System Operators for Electricity}%
\newacronym{ev}{EV}{electric vehicle}%
\newacronym[\glsshortpluralkey={RESs}, \glslongpluralkey={renewable energy sources}]{res}{RES}{renewable energy source}%
\newacronym{it}{IT}{information technology}
\newacronym{snb}{SNB}{Stromnetz Berlin}
\newacronym[\glsshortpluralkey={STs}, \glslongpluralkey={spanning trees}]{st}{ST}{spanning tree}
\newacronym{tso}{TSO}{transmission system operator}
\newacronym{us}{US}{United States}
\newacronym{ub}{UB}{upper bound}
\newacronym{uhv}{UHV}{ultra high voltage}
\makeatletter
\newcommand{\removelatexerror}{\let\@latex@error\@gobble}
\makeatother

\newcommand{\Cflex}{\mathcal{C}^{\uparrow\downarrow}}

\newcommand{\E}{\mathcal{E}}
\newcommand{\efr}{{eft}}
\newcommand{\eto}{{etf}}
\newcommand{\F}{\mathcal{F}}

\newcommand{\fflex}{f^{\uparrow \downarrow}}

\newcommand{\G}{\mathcal{G}}

\newcommand{\I}{\mathcal{I}}

\newcommand{\N}{\mathcal{N}}

\newcommand{\pflex}{p^{\uparrow\downarrow}}
\newcommand{\pflexup}{p^{\uparrow}}
\newcommand{\pflexupopt}{p^{\uparrow\star}}
\newcommand{\pflexdown}{p^{\downarrow}}
\newcommand{\pflexdownopt}{p^{\downarrow\star}}
\newcommand{\qflex}{q^{\uparrow\downarrow}}
\newcommand{\V}{\mathcal{V}}
\newcommand{\Vr}{\mathcal{V}^{\text{ref}}}
\newcommand{\Lo}{\mathcal{L}}

\newcommand{\mvr}{\gls{mv}-Rural\xspace}

\newcommand{\ts}{\textstyle\sum}
\newcommand{\set}[1]{\{#1\}}
\newcommand{\Vflex}{\mathcal{V}^{\uparrow\downarrow}}
\newcommand{\vts}[1]{\lvert #1\rvert}
\newcommand{\Vmin}{\underline{V}}
\newcommand{\Vmax}{\overline{V}}

\begin{document}

\title{Location-Invariant Assessment of Flexibility Potential under Distribution System Reconfiguration
}

\author{\IEEEauthorblockN{Anton Hinneck \IEEEmembership{Member, IEEE}}
\IEEEauthorblockA{
anton.hinneck@stromnetz-berlin.de}
}

\maketitle

\begin{abstract}
    The growing integration of renewable and decentralized generation increases the need for flexibility in distribution systems. This flexibility, typically represented in a PQ capability curve, is constrained by network limits and topology. \Gls{dsr} introduces additional degrees of freedom through switching actions.
This paper proposes an AC-constrained methodology to assess flexibility under network reconfiguration, explicitly considering radial operation. The impact of topology changes on PQ capability curves, which serve as a measure of flexibility potential, is analyzed. To that end, a novel measure called \gls{lifp} is introduced.
Results show that reconfiguration can significantly influence and improve operational flexibility. The approach presented enables transparency for system operators, facilitating improved coordination of flexibility providers.

\end{abstract}


\begin{IEEEdescription}[\IEEEusemathlabelsep\IEEEsetlabelwidth{$~~~~~~~~$}]
\item[$\N$] A graph/power system topology $\set{\E,\V}$
\item[$\V^{(\text{ref})}$] Set of (reference) buses
\item[$\E$] Set of power lines
\item[$\G_{(f)}$] Set of generators (at bus $f$)
\item[$\Lo_f$] Set of loads at bus $f$
\item[$P^{\text{G}}_g$] Real power injection of source $g$
\item[$Q^{\text{G}}_g$] Reactive power injection of source $g$
\item[$p_{\efr}$] Real power flow from bus $f$ to $t$
\item[$q_{\efr}$] Reactive power flow from bus $f$ to $t$
\item[$\pflex$] Real power draw at flex bus
\item[$\qflex$] Reactive power draw at flex bus
\item[$z_{e}$] Switching status of line $\efr$
\item[$V^m_f$] Voltage magnitude at bus $f$
\item[$v^{\text{base}}$] System base voltage
\item[$\Delta \theta_{\eto}$] Voltage angle difference between buses $t$ and $f$ 
\item[$p_l/q_l$] Real/reactive power draw of load $l$
\item[$v^{m,ref}_f$] Measured voltage magnitude at reference bus $f$
\item[$g_{e}/b_{e}$] Conductance/susceptance of line $\efr$
\item[$\Vmin_f/\Vmax_{f}$] Minimum/maximum voltage magnitude at bus $f$
\item[$\overline{s}_{e}$] Maximum power rating of line $\efr$
\item[$z_{e}$] Switching state of line $\efr$
\end{IEEEdescription}

\section{Introduction}
Flexibility means adjusting energy generation or consumption in response to external signals, like prices or activation requests, to support the energy system. It is characterized by factors such as power modulation, duration, response time, and location. As the number of variable \glspl{res} increases, the system requires greater flexibility from both supply and demand to manage unpredictability and maintain stability
\cite{elecpor2014report}.
The rapid growth of technologies such as heat pumps, electric vehicles, and battery storage poses significant challenges to distribution networks due to higher and simultaneous power demand.
Policymakers have long been addressing these challenges alongside efforts from industry. 
Section 13 of the German Energy Industry Act (EnWG) establishes that transmission system operators are responsible for ensuring the security and reliability of the electricity system. 
It defines a hierarchy of measures, i.e., grid-related, market-related, and reserve-based, requiring operators to prioritize cost-efficient solutions when adjusting generation or consumption \cite{EnWG132005}. 
System operators’ authority to intervene in remotely controllable generation facilities has been extended in 2021 to include installations below 100 kW \cite{EnWG13a2021}.
In addition, changes under Section 14a of EnWG in effect since 2024 address the integration of controllable loads and grid connections, driven by the increasing electrification of heating and transport sectors. 
To ensure grid stability and accelerate grid expansion, distribution system operators are granted control mechanisms over installations with a higher rated power than 4.2 kW, which can be complemented by incentives for consumers, such as reduced grid fees \cite{BNetzAEnWG14a2023}. 
Methods for real-time flexibility assessment during operation become key as redispatch measures can exert additional strain on the network - if not assessed properly.
As \glspl{res} are commonly connected at the medium voltage level \cite{buergersolarpark, leimersheim, harzwasserwerke2008}, such assessments are important at the medium voltage level as well.
\newline
Analyses of flexibility potential in active distribution networks are already subject to research \cite{Ageeva2019}.
A commonly used method to assess flexibility is the so-called PQ capability curve, which is generally not convex \cite{Pozo2025}. 
While several methods have been proposed that use power flow approximations for faster computation times \cite{Lopez2021, Paredes2023}, using exact power flow constraints has been shown to be practical \cite{Capitanescu2018, Churkin2023b}.
The interaction of \gls{tso} and \gls{dso} is commonly the subject of research \cite{Silva2018, Capitanescu2018, Churkin2023a} as flexibility unlocked at any voltage level can be transferred across voltage levels and system borders.
Few consider the effects of switching actions. In \cite{Churkin2023b}, the advantage of meshed over radial operation is analyzed.
In this work, and more generally in Germany, both the \gls{mv} and the adjacent \gls{hv} systems are operated by the \gls{dso}. In contrast, the \gls{tso}–\gls{dso} interface connects the \gls{hv} system (typically operated by the \gls{dso}) with the \gls{uhv} system operated by the \gls{tso}.
As previously stated, however, any additional flexibilities unlocked can be provided to the \gls{tso} if system constraints allow.
The contributions in this paper are three-fold:
\begin{enumerate}
\item The impact of \gls{dsr} on flexibility potential in radial distribution systems is analyzed, considering both the \gls{hv} interface and the distribution substation level.
\item A novel metric called \gls{lifp} is introduced to quantify flexibility potential at arbitrary spatial resolutions.
\item The required models using AC power flow are implemented, described, and validated. 
\end{enumerate}

\section{Methodology}
Before the modeling framework and method are introduced in detail, the requirements for the proposed method must be discussed. First, scalable methods for distribution system reconfiguration are required to obtain different topologies $z$ to be analyzed. Moreover, these reconfiguration methods should consider \gls{ac} power flow in order for the results to reliably translate into increases in flexibility. Several methods have recently been proposed in \cite{Hinneck2024} and \cite{Hinneck2025}. The reported resulting topologies in these papers are used as a foundation for flexibility assessment in this paper.
\subsection{Modeling flexibility potential}
\begin{figure*}[ht]
\begin{center}
    \begin{subfigure}[b]{0.32\linewidth}
        \centering
        \includegraphics[width=0.92\linewidth]{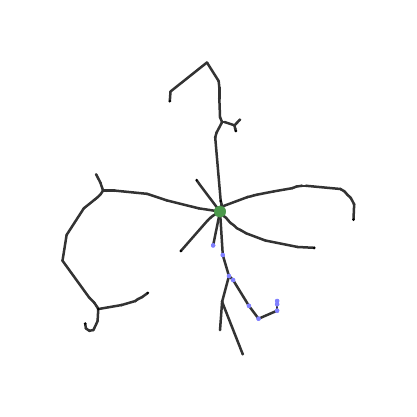}
        \caption{Unfavorable topology}
        \label{fig:topohdsr}
    \end{subfigure}
    \begin{subfigure}[b]{0.32\linewidth}
        \centering
        \includegraphics[width=0.92\linewidth]{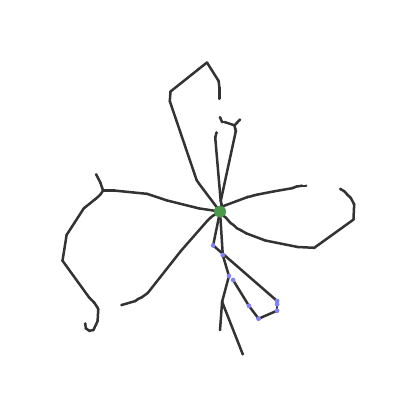}
        \caption{Baseline topology}
        \label{fig:topobase}
    \end{subfigure}
    \begin{subfigure}[b]{0.32\linewidth}
        \centering
        \includegraphics[width=0.92\linewidth]{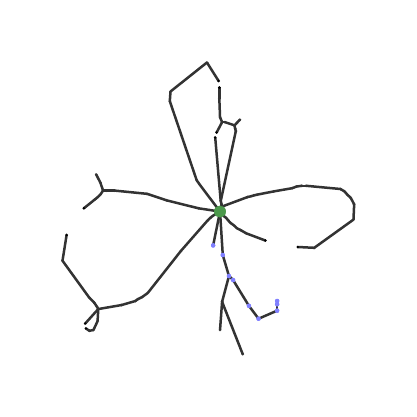}
        \caption{Optimal topology}
        \label{fig:topocdsr}
    \end{subfigure}
    \begin{subfigure}[b]{0.32\linewidth}
        \centering
        \includegraphics[width=0.98\linewidth]{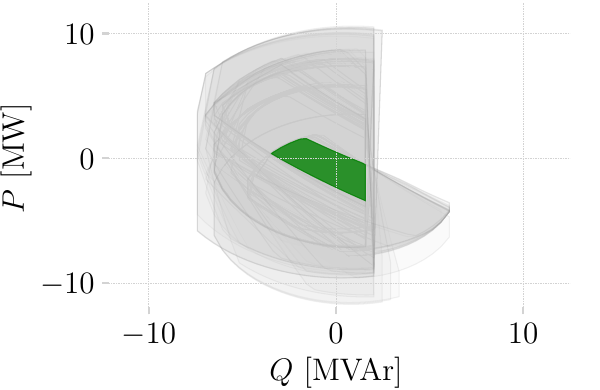}
        \caption{Unfavorable topology}
        \label{fig:pqhdsr}
    \end{subfigure}
    \begin{subfigure}[b]{0.32\linewidth}
        \centering
        \includegraphics[width=0.98\linewidth]{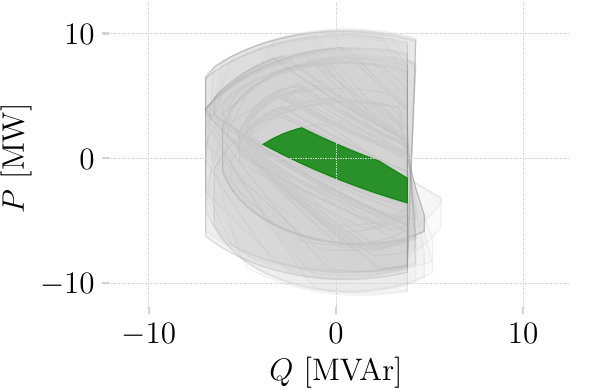}
        \caption{Baseline topology}
        \label{fig:pqbase}
    \end{subfigure}
    \begin{subfigure}[b]{0.32\linewidth}
        \centering
        \includegraphics[width=0.98\linewidth]{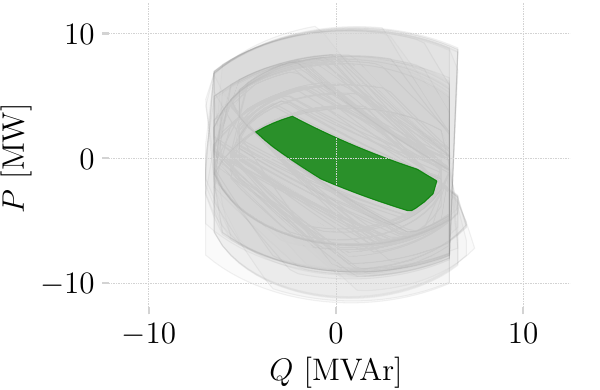}
        \caption{Optimal topology}
        \label{fig:pqcdsr}
    \end{subfigure}
    \caption{This figure displays the \glspl{lifp} $\hat{\F}_{\Vflex}$ for different operational topologies in green. Gray surfaces display the \glspl{nfp}. The topologies are displayed above for easy reference. The topologies were determined optimizing for minimal losses which produces balanced system utilization \cite{Hinneck2024}. This balanced utilization also shows to maximize $\hat{\F}_{\Vflex}$ implicitly, unlocking flexibility in the system.}
    \label{fig:alldss}
\end{center}
\end{figure*}
\subsubsection{Nodal flexibility potential}
The modeling framework proposed in this paper aims to assess host capacity as realistically as possible. As such, \gls{ac} power flow constraints are used. The real power balance at every bus is enforced by constraints \eqref{acflex:mcr}, whereas \eqref{acflex:mcr} enforces the reactive power balance. The bus subject to assessment is denoted as $\fflex$. The variable $\pflex$ is only considered in the real power power balance constraint that is currently subject to assessment, i.e., $f=\fflex$.
$\qflex$ is a parameter in the reactive power flow constraints used to map the feasibility potential in the PQ plane. 
The constraints \eqref{acflex:pfr} and \eqref{acflex:qfr} denote \gls{ac} line flow constraints of real and reactive power flowing into and out of transmission lines.
Constraint \eqref{acflex:ref} sets the voltage magnitude at the reference bus.
Most importantly for the derivation of flexibility potential, operational constraints are enforced. 
Constraints \eqref{acflex:voltagelimits} ensure that the voltage magnitude does not leave the required interval.
In \eqref{acflex:powerlimits}, thermal ratings of transmission lines are enforced. 
These constraints \eqref{acflex:voltagelimits} \& \eqref{acflex:powerlimits} ultimately limit flexibility.
\begin{subequations}
\allowdisplaybreaks
\label{mod:rdsr}
\begin{alignat}{2}
    \ts_{g\in\G_{f}}p^{\text{G}}_{g} 
    & = \ts_{t\in\E_f} p_{eft} + \ts_{t\in\E_f} p_{etf} \nonumber\\
    &\quad + \ts_{l\in\Lo_{f}}p_{l} - \pflex\mathbf{1}_{\{f = \fflex\}}, 
    && \forall~ f\in\V \label{acflex:mcr}\\
    \ts_{g\in\G_{f}}q^{\text{G}}_{g}
    & = \ts_{t\in\E_f} q_{eft} + \ts_{t\in\E_f} q_{etf} \nonumber\\
    &\quad + \ts_{l\in\Lo_{f}}q_{l} - \qflex\mathbf{1}_{\{f = \fflex\}}, 
    && \forall~ f\in\V \label{acflex:mci}\\
    p_{eft} 
    & = {\color{white} -}\big({V^m_f}^2 (g_{e}+g^{sh}_{e}) \nonumber\\
    &\quad - V^m_{f}V^m_{t}g_{e}\cos(\Delta \theta_{eft}) \nonumber\\
    &\quad - V^m_{f}V^m_{t}b_{e} \sin(\Delta \theta_{eft})\big)z_{e}, 
    && \forall~ eft,etf\in\E \label{acflex:pfr}\\
    q_{eft} 
    & = -\big({V^m_f}^2 (b_{e}+b^{sh}_{e}) \nonumber\\
    &\quad + V^m_{f}V^m_{t}b_{e}\cos(\Delta \theta_{eft}) \nonumber\\
    &\quad - V^m_{f}V^m_{t}g_{e} \sin(\Delta \theta_{eft})\big)z_{e}, 
    && \forall~ eft,etf\in\E \label{acflex:qfr}\\
    V^m_{f} 
    & = v^{m,ref}_{f},
    && \forall~ f\in\Vr \label{acflex:ref}\\
    \underline{V}^m_{f} 
    & \leq V^m_{f} \leq \overline{V}^{m}_{f},
    && \forall~ f\in\V \label{acflex:voltagelimits}\\
    \overline{s}_{e}^2 
    & \geq p_{eft}^2 + q_{eft}^2,
    && \forall~{eft}\in\E \label{acflex:powerlimits}
\end{alignat}
\end{subequations}
While constraints \eqref{acflex:mcr}-\eqref{acflex:powerlimits} already define a feasible set, the notation is further generalized for easy reference.
Let $\F_{\fflex}$ denote the \gls{nfp}. It is now defined as;
\begin{equation}
    \F_{\fflex}:=\{ \pflex,p^{G},q^{G},p,q,\theta, V^m \mid \eqref{acflex:mcr}-\eqref{acflex:powerlimits} \}.
\end{equation}
To represent $\F_{\fflex}$ in a computationally efficient and processable form, the set's boundary $\partial\F_{\fflex}$ is commonly sampled as is in the following.
Hence, $\partial\F_{\fflex}$ a piecewise polygonal approximation is derived.
This is achieved by determining tuples $(\pflexupopt_i,\qflex_i)$ and $(\pflexdownopt_i,\qflex_i)$ solving $2\vts{\I}$ optimization problems where $\{\qflex_1,\dots,\qflex_{\vts{\I}}\}$ are parameters, which are uniformly spaced over the interval $[\qflex_1, \qflex_{\vts{\I}}]$ in this work. To sample the boundary, optimization problems are defined next to determine objective values $\pflexupopt_i, \pflexdownopt_i~\forall i\in\I$. To construct \textsc{AC-Flex-Max} $\pflex$ and $\qflex$ are substituted by $\pflexup_i$ and $\qflex_i$ respectively. This particularly applies to \eqref{acflex:mcr} and \eqref{acflex:mci} and results in optimization problem \eqref{mod:flexmax}.
\begin{equation}
\label{mod:flexmax}
\underset{\qflex_i,\fflex}{\textsc{AC-Flex-Max}}:\quad
\left\{
\begin{aligned}
\max_{\pflexup_i,p^{G},q^{G},p,q,\theta, V^m} \quad & \pflexup_i\\
\text{s.t.} \quad  \eqref{acflex:mcr}-&\eqref{acflex:powerlimits}
\end{aligned}
\right.
\end{equation}
Analogously, to construct \textsc{AC-Flex-Min}, $\pflex$ and $\qflex$ are substituted by $\pflexdown_i$ and $\qflex_i$ respectively. This results in minimization problem \eqref{mod:flexmin}. 
\begin{equation}
\label{mod:flexmin}
\underset{\qflex_i,\fflex}{\textsc{AC-Flex-Min}}:\quad
\left\{
\begin{aligned}
\min_{\pflexdown_i,p^{G},q^{G},p,q,\theta, V^m} \quad & \pflexdown_i\\
\text{s.t.} \quad  \eqref{acflex:mcr}-&\eqref{acflex:powerlimits}
\end{aligned}
\right.
\end{equation}
To analyze the flexibility potential of a bus $\fflex$, \textsc{AC-Flex-Min} and \textsc{AC-Flex-Max} must be solved for all tuples $\fflex\times\{\qflex_1,\dots,\qflex_{\vts{\I}}\}$.
To estimate the \gls{nfp} following polygonal curve is generated;
\begin{align}
    \Cflex_{\fflex} := \{(\pflexupopt_1(\fflex,\qflex_1),\qflex_1),&\dots,(\pflexupopt_{\vts{\I}}(\fflex,\qflex_{\vts{\I}}),\qflex_{\vts{\I}}),\nonumber\\
    (\pflexdownopt_{\vts{\I}}(\fflex,\qflex_{\vts{\I}}),\qflex_{\vts{\I}}),
    &\dots,(\pflexdownopt_1(\fflex,\qflex_1),\qflex_1),\nonumber\\
    (\pflexupopt_1(\fflex,\qflex_1),\qflex_1)\}\label{eq:Cconstruction}
\end{align}
The final set $\Cflex_{\fflex}$ may be smaller than indicated in \eqref{eq:Cconstruction}, as certain instances of \eqref{mod:flexmin} and \eqref{mod:flexmax} may not be solvable and the respective tuples are not included.
One can clearly observe, however, that the computational cost scales linearly in~$\mathcal{I}$. 
However, there is potential for parallelization, as sampling different points can be performed independently.
The polygonal approximation of the \gls{nfp} $\F_{\fflex}$ of bus $\fflex$ is defined in \eqref{def:nodalflexapprox}. In the following, while symbol $\tilde{F}_{\fflex}$ will be used, the set will likewise be simply referred to as \gls{nfp} to ensure conciseness.
\begin{equation}
    \label{def:nodalflexapprox}
    \tilde{\F}_{\fflex}:=\Cflex_{\fflex}\cup\mathrm{interior}(\Cflex_{\fflex})
\end{equation}
The larger the set $\vts{\I}$, the more points of the feasibility potential are sampled, and the more accurate $\tilde{\F}_{\fflex}$.
Sampling more points, however, also increases computational complexity, which requires a proper balance to be struck.
\subsubsection{Location-invariant flexibility potential}
Having determined the \gls{nfp} at multiple buses $\fflex_j\in\Vflex$, an aggregated potential of that set can be assessed. This set is called \gls{lifp} and is defined as
\begin{equation}
    \hat{\F}_{\Vflex} := \textstyle\bigcap_{\fflex\in\Vflex}\tilde{\F}_{\fflex_j}.
\end{equation}
It is location-invariant—or robust, so to speak—with respect to all busses in aggregation zone $\Vflex$.
Here, the choice $\vts{\I}$ becomes even more important when considering computational tractability, as a problem instance must be solved for every tuple $\{\fflex_1,\dots,\fflex_{\vts{\Vflex}}\}\times\{\qflex_1,\dots,\qflex_{\vts{\I}}\}$, i.e., $\vts{\I}\vts{\Vflex}$ tuples in total.
The full procedure of computing the set is illustrated in Algorithm \ref{alg:flex}.
\begin{algorithm}
\DontPrintSemicolon
\SetAlgoNoEnd
\LinesNumbered
\caption{Construction of $\hat{\F}_{\Vflex}$}
\label{alg:flex}
\KwIn{$\Cflex$, $\{\qflex_1,\dots,\qflex_{\vts{\I}}\}$}
$\hat{\F}_{\Vflex} = \mathbb{R}^2$\;
\ForEach{$\fflex \in \Vflex$}{
    $\mathcal{C}^{\uparrow}_{\fflex} = \emptyset$, $\mathcal{C}^{\downarrow}_{\fflex} = \emptyset$\;
    \For{$i \in\I$}{
        $p^{\uparrow/\downarrow\star}_i \gets \text{solve}\Big(\underset{\qflex_i,\fflex}{\textsc{AC-Flex-Max/Min}}\Big)$\;
        \If{Solution $p^{\uparrow/\downarrow\star}_i$ was found}{
            $\mathcal{C}^{\uparrow/\downarrow}_{\fflex} \gets (p^{\uparrow/\downarrow\star}_i(\fflex,\qflex_i),\qflex_i)$\;
            {\color{lightgray}\tcc{push last/first}}
        }
    }
    $\tilde{\F}_{\fflex} = \text{construct}(\mathcal{C}^{\uparrow}_{\fflex}, \mathcal{C}^{\downarrow}_{\fflex})$\;
    {\color{lightgray}\tcc{based on \eqref{eq:Cconstruction} and \eqref{def:nodalflexapprox}}}
    $\hat{\F}_{\Vflex} \gets \hat{\F}_{\Vflex}\cap\tilde{\F}_{\fflex}$
}
\Return{$\hat{\F}_{\Vflex}$}\;
\end{algorithm}
Note, that lines 3-8 summarize the procedure applied to obtain \glspl{nfp}, i.e., $\tilde{\F}_{\fflex}$.
\subsection{\glspl{lifp} in system operation}
\begin{figure*}[ht]
\begin{center}
    \begin{subfigure}[b]{0.32\linewidth}
        \centering
        \includegraphics[width=0.98\linewidth]{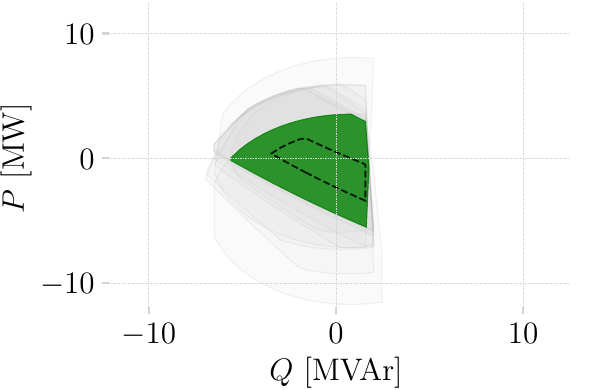}
        \caption{Unfavorable topology}
        \label{fig:pqhdsrsubset}
    \end{subfigure}
    \begin{subfigure}[b]{0.32\linewidth}
        \centering
        \includegraphics[width=0.98\linewidth]{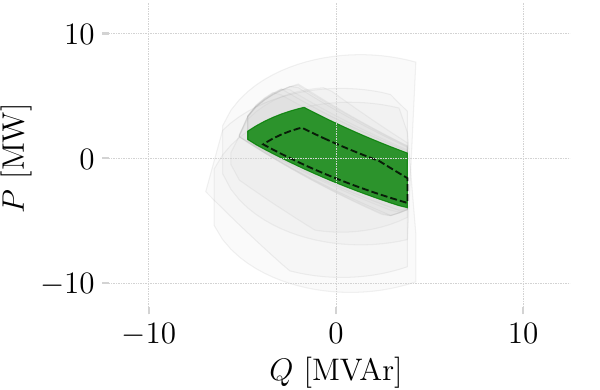}
        \caption{Baseline topology}
        \label{fig:pqbasesubset}
    \end{subfigure}
    \begin{subfigure}[b]{0.32\linewidth}
        \centering
        \includegraphics[width=0.98\linewidth]{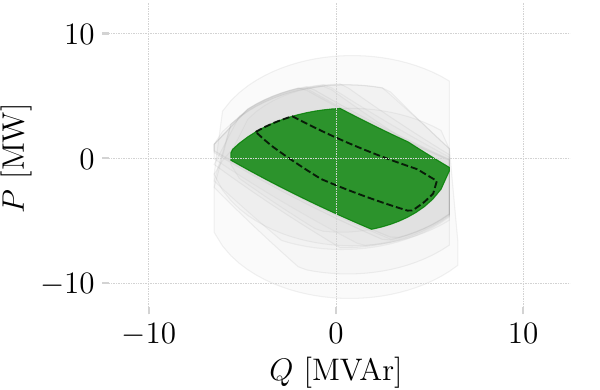}
        \caption{Optimal topology}
        \label{fig:pqcdsrsubset}
    \end{subfigure}
    \caption{This figure displays \glspl{lifp} for the open ring, highlighted in Figure \ref{fig:alldss}, in green. The \glspl{lifp} for the entire system are inscribed as dotted black curves for reference. Gray surfaces display \glspl{nfp}.}
    \label{fig:subset}
\end{center}
\end{figure*}
Based on the size of aggregated regions, i.e., $\vts{\Vflex}$, \glspl{lifp} can be computed continuously. This provides system operators with a real-time overview of the maximum admissible flexibility within a region, avoiding the need for per-bus validation. Such continuous assessment is valuable, as operational actions may adversely affect available flexibility. In particular, switching actions can improve voltage profiles and load flows, but may also reduce the capability of generators and controllable devices to provide redispatch or ancillary services, as demonstrated in Section~\ref{sec:results}. The metric $\hat{\F}_{\Vflex}$ quantifies the worst-case capability within a region—potentially the entire grid—across all flexibility providers.
\newline
It is important to note that \glspl{lifp} and \glspl{nfp} are closely related. One can observe that both measures provide the same PQ capability curve if $\Vflex = \{\fflex\}$.

\section{Results}
\label{sec:results}
The procedure is validated on a real 95-bus distribution system included in the SimBench \cite{Meinecke2020} data set called \mvr.
The base voltage is $v^{\text{base}}=20~\text{kV}$. 
The was converted from PandaPower to a MATPOWER case file. All its properties are stated after conversion.
The total load summed over all buses amounts to 17.26 MW and 6.82 MVAr. The total generation of \glspl{res} amounts to 25.57 MW. Hence, excluding all transmission losses, 8.31 MW is exported to the \gls{hv} grid in this scenario with high \gls{res} infeed. This system has the structure of a typical \gls{mv} system in Germany. It is comprised of open rings, which results in a radial operational topology. All models were solved using JuMP, the Julia programming language, and IPOPT.
\subsection{Determining optimal topologies}
\begin{figure}[h!]
    \centering
    \includegraphics[width=0.98\linewidth]{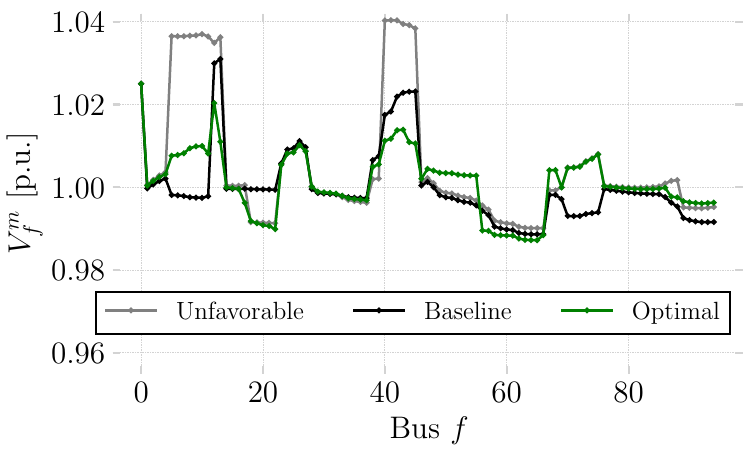}
    \caption{This figure displays the location-invariant hosting capacity for flexibility services for different nominal topologies.}
    \label{fig:voltage}
\end{figure}

Different operational topologies are compared for this test case.
Recent work has proposed and tested heuristics \cite{Hinneck2024} and exhaustive methods \cite{Hinneck2025} for distribution system reconfiguration.
The identified topologies are used for further analysis in this study.
The \textit{baseline} topology is included in the test case. The \textit{unfavorable} topology is generated using a restricted heuristic problem instance from \cite{Hinneck2024} and the optimal topology using a model that considers the complete solution space \cite{Hinneck2025}. The resulting topologies are displayed in Figures \ref{fig:topohdsr}-\ref{fig:topocdsr}. The resulting voltage profiles are illustrated in Figure \ref{fig:voltage}.
\begin{figure*}[ht]
\begin{center}
    \begin{subfigure}[b]{0.31\linewidth}
        \centering
        \includegraphics[width=0.98\linewidth]{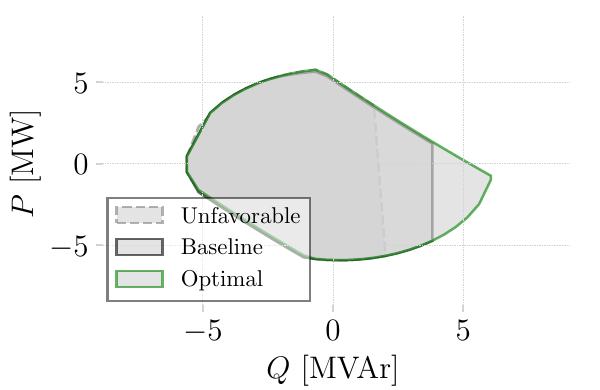}
        \caption{Bus 25}
        \label{fig:bus25}
    \end{subfigure}
    \begin{subfigure}[b]{0.31\linewidth}
        \centering
        \includegraphics[width=0.98\linewidth]{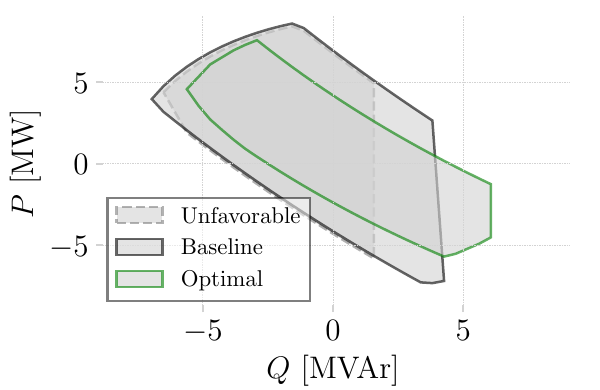}
        \caption{Bus 56}
        \label{fig:bus56}
    \end{subfigure}
    \begin{subfigure}[b]{0.31\linewidth}
        \centering
        \includegraphics[width=0.92\linewidth]{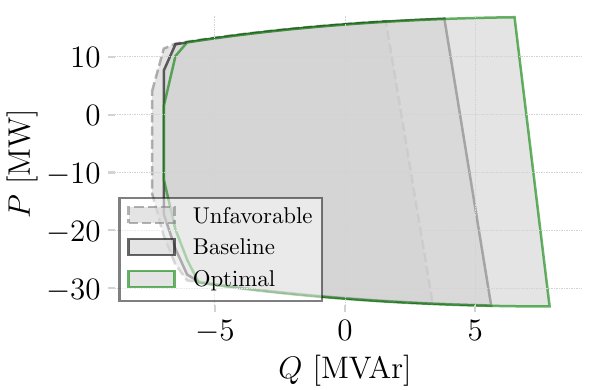}
        \caption{HV/MV interface}
        \label{fig:interface}
    \end{subfigure}
    \caption{This figure displays the \glspl{nfp} at different buses for different nominal topologies.}
    \label{fig:nodalsensitivities}
\end{center}
\end{figure*}
It can be observed that in the unfavorable topological configuration $V^m$ at several buses is sharply raised due to decentral \gls{res} infeed and close to the limit of $0.95~\text{p.u.} \leq V^m\leq 1.05~\text{p.u.}$ The baseline and optimal topology effect a more level voltage profile. This is a consequence of the objective function used in \cite{Hinneck2024,Hinneck2025}.
\subsection{Sensitivity of \glspl{lifp}}
%
The \gls{lifp} is highly sensitive to topological actions, as shown in Fig.~\ref{fig:alldss}. Certain topologies unlock significant flexibility potential. For the unfavorable topology, the upper voltage limit in \eqref{acflex:powerlimits} is nearly reached, restricting capacitive reactive power provision. Improved topologies alleviate this constraint by reducing maximum voltages and increasing operational margins.
For a quantitative comparison, the normalized areas of the PQ capability curves are summarized in Table~\ref{tab:pqfes}. The optimal topology increases the PQ capability area by 173\% compared to the least favorable configuration and by 66\% relative to baseline. This is illustrated by the green areas in Figures \ref{fig:pqhdsr}-\ref{fig:pqcdsr} or dashed areas in Figures \ref{fig:pqhdsrsubset}-\ref{fig:pqcdsrsubset} respectively.
\begin{table}[h]
\centering
\caption{Normalized PQ capability areas measuring operational flexibility unlocked by different operational topologies}
\label{tab:pqfes}
\begin{tabular}{lccc}
\hline
Topology & Norm. PQ capability & Improvement over previous [\%] \\
\hline
Unfavorable & 1.00 & -- \\
Baseline & 1.64 & 64 \\
Optimal & 2.73 & 66 \\
\hline
\end{tabular}
\label{tab:pq_area_comparison}
\end{table}
\newline
The potential is sensitive to the locations considered, i.e., $\Vflex$. The effects of this are illustrated in Figure \ref{fig:subset}. Here, $\Vflex$ only includes buses of a single open \gls{mv} ring, the distribution substations of which are highlighted in blue in Figure \ref{fig:alldss}. The following behavior is expected: For equal topologies, \gls{lifp}-1 constructed for a region $\Vflex_1$ must be greater or equal to \gls{lifp}-2, constructed for region $\Vflex_2$, if $\Vflex_1\subseteq\Vflex_2$.
This holds in this case study as set set ob blue buses is clearly a subset of all buses. For easy reference, the global \gls{lifp} in inscribed in the \gls{lifp} of the open ring. This illustrates that the concept of \gls{lifp} allows for arbitrary assessment and aggregation. Moreover, \gls{dsr}, as conducted in \cite{Hinneck2024,Hinneck2025}, shows significant positive effects on available flexibility.

\subsection{Nodal flexibility potentials}
Three different effects are presented in the following. First, applying a topology that levels voltage profiles unlocks flexibility not just overall but also locally. This can be clearly seen at the example of the \gls{nfp} at bus 25 in Figure \ref{fig:bus25}.
Counterintuitively, however, an overall optimal topology can reduce flexibility locally to the benefit of the entire system. This can be observed in Figure \ref{fig:bus56}, especially when comparing the \gls{nfp} for the baseline and optimal topologies.
Here, not just voltage limits but power limits lead to differing \glspl{nfp}.
Lastly, the effect on the \gls{hv}/\gls{mv} interface is analyzed, specifically focusing on the \gls{nfp} at the \gls{hv} busbar. 
One can again observe, in Figure \ref{fig:interface}, that reconfiguration extends the \gls{nfp} significantly.
This again shows a positive impact of \gls{dsr} on flexibility provision.

\section{Conclusion}
\label{sec:conclusion}
A framework for assessing flexibility potential in arbitrary spatial resolution is presented. It is applied to analyze the effects of \gls{dsr} on PQ capability in the network.
Results indicate that network topology significantly affects flexibility. Reconfiguration can increase feasible operating regions by alleviating constraints, while effects remain location-dependent and may introduce trade-offs between local and system-wide flexibility.
The methodology enables spatially resolved flexibility assessments and supports operational analyses under varying network topologies.
Improving computational performance, and the incorporation into market designs are relevant directions for future research.

\bibliographystyle{IEEEtran}
\bibliography{bib}

\end{document}